\documentclass[10pt]{iopart}

\usepackage{amstext}
\usepackage{graphicx}

\def\bea{\begin{eqnarray}}
\def\eea{\end{eqnarray}}

\begin{document}

\title{Scale-local dimensions of strange attractors}

\author{J G Reid and T A Trainor}

\address{CENPA, Box 354290, University of Washington, Seattle, Washington 98195-4290}

\ead{trainor@hausdorf.npl.washington.edu}

\begin{abstract}
We compare limit-based and scale-local dimensions of complex distributions, particularly for a strange attractor of the H\'{e}non map.  Scale-local dimensions as distributions on scale are seen to exhibit a wealth of detail.  Limit-based dimensions are shown to be averages of scale-local dimensions, in principle over a semi-infinite scale interval. We identify some critical questions of definition for practical dimension analysis of arbitrary distributions on bounded scale intervals. 
\vskip .2in
keywords: scale, dimension, information, correlation integral, H\'enon map
\end{abstract}


\submitto{\JPA}

\maketitle


\section{Introduction}

In the past two decades, improved understanding of dimension has been stimulated by studies of nonlinear dynamical systems and  their strange attractors. Canonical dimensions used to characterize strange attractors have been defined in terms of zero-scale limits. These limit-based dimensions are inferred by varying a partition scale and extrapolating results to an invariant limit. In this paper we consider {\em scale-local} R\'{e}nyi dimensions of a strange attractor of the H\'enon map as explicit functions of scale.  We demonstrate that approximations to the zero-scale limit by extrapolation represent scale averages of scale-local dimension. We also contrast dimension defined as a topological invariant with dimension based on numerical analysis of arbitrary distributions and the problem of estimation based on computationally-accessible scale intervals, for which extrapolations based on {\em a priori} considerations are not relevant.

In this paper we review limit-based and scale-local dimension definitions, identify limit-based dimensions as scale averages of scale-local dimensions, survey previous limit-based dimension analyses of a H\'{e}non attractor, present a scale-local dimension analysis of this attractor, and examine the general structure of scale-local and running-average dimension distributions and their monotonicity with R\'{e}nyi index $q$. We conclude by examining the status of limit-based dimensions as topological invariants and their relation to scale-local dimension, information and dimension transport.

\section{Limit-based Dimensions}

Dimensionality is a fundamental property of distributions characterizing their correlation structure. Modern dimension theory is based on the work of Carath\'{e}odory and Hausdorf \cite{pesin,haus,czyz}. Dimension is there defined in terms of asymptotic limits of set measures depending on partition scale. Limit-based dimensions are invariant properties of distributions also invariant under certain transformations.

Hausdorff-Besicovitch dimension \cite{falc} is based on the set measure
\begin{eqnarray}
H^{\kappa}(E,e) = \inf \left( \sum^\infty_{i=1} h(B_i) :~ B_i \subseteq 
\underline{E},~ e_i \leq e \right), \label{hausm}
\end{eqnarray}
where $\{B_i\}$ is a partition of the embedding space, $\underline{E}$ is the {\em support} of set $E$ and $e_i$ is the size of the $i^{th}$ partition  element. The characteristic scale $e$ of a Hausdorff partition is an {\em upper limit} on partition-element size. If the point measure or dimension function is assumed to have a power-law form $h(B_i) \propto e_i^{\kappa}$ with $\kappa$ arbitrary,  then Hausdorff-Besicovitch dimension $d^{\text{h}}$ is the value of $\kappa$ for which $\lim_{e \rightarrow 0} H^{\kappa}(E,e)$ remains finite, $\in (0,\infty)$.

Box-counting dimension is based on partition elements (boxes) which have a common size (in contrast to Hausdorff partitions). The number $M(E,e)$ of boxes of size $e$ required to cover a set $E$ is
\begin{eqnarray}
M(E,e) = \sum_i : B_i(e) \subseteq \underline{E},
\end{eqnarray}
For self-similar sets $M(E,e)$ is expected to follow a power law on scale: $M(E,e) \propto e^{-d^{\text{b}}}$, where $d^{\text{b}}$ is the fractal or box-counting dimension, in which case
\begin{eqnarray} \label{boxdim}
d^{\text{b}} &\equiv& \lim_{e \rightarrow 0} {\log M(E,e) \over \log L/e}. 
\end{eqnarray}
with $L$ the boundary size of $E$. By analogy one can define a box-counting set measure \cite{rudo}
\begin{eqnarray}
K^{\kappa}(E,e) &=& \sum_i e^{\kappa} : B_i  \subseteq \underline{E}, e_i = e \nonumber \\ \label{boxm}
&=& M(E,e) \cdot e^{\kappa}. \label{hausbox}
\end{eqnarray}
The limit $\lim_{e \rightarrow 0} K^{\kappa}(E,e)$ should remain finite for $\kappa$ = $d^{\text{b}}$. Box-counting dimension can be evaluated with the computationally convenient Eq.~(\ref{boxdim}), whereas Hausdorff set measure, an {\em infinite} sum over partition elements, cannot be factored as in the second line of Eq.~(\ref{hausbox}).

Hausdorff and box-counting dimensions were initially applied to strange attractors with the goal of better understanding the nonlinear processes which define such distributions. The dimension concept was later expanded to information, correlation and higher R\'{e}nyi dimensions \cite{gollub}, reflecting multipoint ($q$-point) correlations of measure distributions. These dimensions are also conventionally defined in terms of zero-scale limits.

Box-counting and Hausdorff dimensions correspond to $q=0$. Information dimension $d^{\text{i}}$ is defined $via$ application of l'H\^{o}pital's rule in the limit $q \rightarrow 1$. Correlation dimension $d^{\text{c}}$ \cite{grassprl} measures two-point ($q = 2$) correlations as defined by the correlation integral
\bea
C_2(E,e) &=& {1 \over N^2(E)} \sum_{i, j=1}^N \theta(e-|x_i-x_j|) \\ \nonumber
&\simeq& \sum_{i=1}^{M(E,e)} n_i^{[2]}(E,e)/N^{[2]}(E) ,
\eea
with $C_2(E,e) \propto e^{d^{\text{c}}}$ as $e \rightarrow 0$ in the case of a fractal. In general, $d^{\text{h}} \simeq d^{\text{b}} \equiv d^{\text{r}}_0$, $d^{\text{i}} \equiv d^{\text{r}}_1$ and $d^{\text{c}} \equiv d^{\text{r}}_2$, with the R\'{e}nyi dimensions \cite{renyi} defined by $d^{\text{r}}_q(E) \equiv \lim_{e \rightarrow 0} \left\{S_q(E,e) / \log(L/e) \right\} $ and
\begin{eqnarray} \label{renent}
S_q(E,e) &\equiv&  \frac{1}{1-q} \log C_q(E,e) \\ \nonumber
&\simeq& \frac{1}{1-q} \log\left[\sum_{i=1}^{M(E,e)} p_i^q(E,e)\right] ,
\end{eqnarray}
where we approximate the rank-$q$ normalized correlation integral by $C_q \approx \sum_i n_i^{[q]} / N^{[q]}$ \cite{scanew}, $n^{[q]} = n! / (n-q)!$, $n_i$ is the normalized contents (point count or measure total) of the $i^{th}$ partition element, $p_i^q \approx \langle n_i^{[q]}/N^q \rangle \approx \langle n_i/N \rangle^q$, $L$ is the boundary size and $N$ is the total point count or measure in $E$.

\section{Scale-local Dimensions}

For the analysis of arbitrary distributions over scale intervals limited by computational or experimental accessibility we must understand dimension in a larger context without assumptions of recursiveness and extrapolation to asymptotic limits (generalizing beyond distributions defined by recursive mappings or iterated part removal). Within bounded scale intervals distinctions are required which are not relevant for limit-based dimensions.

Scale-local dimension, in contrast to limit-based dimension, is an explicit function of scale relating {\em relative change of a measure to relative change of scale} at arbitrary scale and space points 
\begin{eqnarray} 
d(\mu,e,\delta \log e) \equiv \frac{1}{\mu(\log e)}\, \frac{\mu(\log e+0.5\,\delta \log e) - \mu(\log e-0.5\,\delta \log e)}{\delta \log e}. 
\end{eqnarray} 
represented compactly by $d(\mu,e,\delta \log e) \rightarrow { \partial \log \mu / \partial \log e}$. The dimension distribution on scale depends on scale resolution $\delta \log e$. The convenient differential notation should not imply an asymptotic resolution limit ($\delta \log e$ will be suppressed notationally but must be specified for numerical analysis). In contrast to limit-based dimensions no {\em a priori} assumption is made about scale dependence or asymptotic values. 

Dimension $\tau_q(E,e)$  relates relative change in a correlation integral to relative change in scale
\begin{eqnarray}
\tau_q(E,e) &=& \partial \log C_q(E,e)/\partial \log e .
\end{eqnarray}
$\tau_q$ represents the {\em projection} of a $q\cdot d$-dimensional space onto a $(q-1)\cdot d$-dimensional subspace or hyperplane perpendicular to its main diagonal, the domain of the correlation integral $C_q(e)$. $d_q(E,e)$ is the corresponding rank-q dimension of set $E$ in the $d$-dimensional primary space defined by \cite{scanew}
\begin{eqnarray}
 (q-1)\cdot d_q(E,e) &=& \tau_q(E,e).
\end{eqnarray}
Scale-local dimensions are expressed in terms of R\'{e}nyi entropies  Eq.~(\ref{renent}) as
\begin{eqnarray} \label{renyidim}
d_q(E,e)  &=& {1 \over q-1} {\partial \log C_q(E,e) \over \partial \log e} \nonumber \\
&=& \partial S_q(E,e)/\partial \log 1/e \nonumber \\
&\simeq& {1 \over q-1} {\partial \log \sum_{i=1}^{M(E,e)} p_i^q(E,e) \over \partial \log e}. 
\end{eqnarray}

\section{Relating Limit-based and Scale-local Dimensions} \label{relate}

Entropy $S_q(E,e,L)$ depends on {\em scale intervals}: $S_0(E,e,L) = \log M(E,e,L)$ is determined by the number of bins of size $e$ within a distribution boundary of size $L$. R\'{e}nyi entropies are expressed as definite integrals of scale-local dimensions by inverting Eq.~(\ref{renyidim})
\begin{eqnarray} \label{mq}
S_q(E,e,L) &=&  \int_{\log e}^{\log L}d_q(E,e')d\log e'.
\end{eqnarray}
Limit-based R\'{e}nyi dimensions in the form
\bea \label{dlim}
d_q^{\text{r}}(E) &\equiv& \lim_{e \rightarrow 0} \frac{S_q(E,e,L)}{\log L/e} \\ \nonumber
&=& \lim_{e \rightarrow 0} \frac{\int_{\log e}^{\log L}d_q(E,e')d\log e'}{\int_{\log e}^{\log L}d\log e'} \\ \nonumber
&\equiv& \lim_{e \rightarrow 0} \overline{d_q(E,e,L)}.
\eea
are thereby interpreted as {\em mean values} of scale-local dimensions averaged (in principle) over a `semi-infinite' scale interval bounded above by the boundary size of the distribution and below (in practice) by finite computing power.

If distribution $E$ is described by a single dimension value over the averaging interval the distribution on dimension is then a delta function (basis for idealized arguments) and the mean obviously takes this value. More generally, limit-based dimension is the mean of a finite-width distribution on dimension. The width and skewness of the distribution may also be of interest for characterizing attractors. The distribution may even be multimodal (for more complex scale dependencies) or have no central tendency. 

The mean value defined as a zero-scale limit in Eq.~(\ref{dlim}) can be generalized to an explicit function of scale interval: $\overline{d_q(E,e_1,e_2)}$. Scale interval $[\log e_1,\log e_2]$ then ranges from the limit-based $(-\infty,\log L]$ to the scale-local $[\log e - 0.5 \, \delta \log e,\log e + 0.5 \,\delta \log e]$ and may be chosen to minimize boundary-scale and void-bin biases in a scale-local dimension distribution. Variation of $\overline{d_q(E,e_1,e_2)}$ with decreasing lower bound $e_1$ may give a misleading impression of convergence to $\lim_{e \rightarrow 0} d_q(E,e)$. Integration over an increasing scale interval attenuates scale-local variation in proportion to the averaging interval, suggesting convergence to a zero-scale dimension value (if it exists) which is not necessarily the limit of the mean.

\section{The H\'enon Map}

For illustration we consider limit-based and scale-local dimensions of a well-known strange attractor of the H\'{e}non map \cite{henon}. 
\begin{eqnarray}
x&\mapsto&a+by-x^2 \\ \nonumber
y&\mapsto&x.
\end{eqnarray}
The attractor we analyze corresponds to parameter values $a=1.4$ and $b=0.3$, whose properties have been extensively described in the literature, and falls in the bounded region $x,y \in [-1.8,1.8]$.  Orbits are either trapped within this region or unbounded \cite{BRHunt}. This attractor is useful because it demonstrates central tendency (on dimension) in the $d_q(E,e)$ distributions on scale, and the $rms$ variations of the distributions are significant: $\sim 5$\% of mean values as indicated in Fig.~\ref{henplotsa1}. 

A finite-length mapping generates a sample of the attractor, not the attractor itself (the parent distribution for the sample). For each mapping we choose an initial random seed $(x_0,y_0)$ and discard iterations until we sample points on the attractor itself.  Increased sample number (larger point density) reduces the small-scale limit of the dimension distribution.  Multiple sampling with randomly-chosen seeds indicates which dimension characteristics are independent of initial conditions.

\section{Limit-based H\'{e}non Dimensions} \label{limbase}

The limit-based dimension $d^{\text{b}}(E)$ is defined in principle by Eqs.~(\ref{boxdim},\ref{classdim}).  
\bea \label{classdim}
d^{\text{b}} \equiv \lim_{e \rightarrow 0} \frac{\log M(E,e)}{\log L/e}.
\eea
The boundary size $L$ is often not specified, assumed to be unity or used as a free fit parameter. In the limit, the choice of $L$ would not matter. For practical analysis on a bounded scale interval the choice of $L$ may be a critical bias source for dimension estimates.

Value $d^{\text{b}} = 1.261 \pm .003$ \cite{russell} was determined by a linear least-squares fit to $\log M(E,e)$ {\it vs.} $\log e$ over a bounded scale interval, with $L$ a free fit parameter. $\chi^2$ minimization determines an effective dimension definition (for fit-optimized $L$)
\bea \label{avdim}
d^{\text{b}} & \approx & \frac{\overline{\log M(e)}}{\overline{\log L / e}} = \frac{\sum_i \log M(e_i) \cdot \log L / e_i }{\sum_i \{\log L / e_i \}^2 }, 
\eea
a ratio of weighted scale averages favoring smaller scales. Since each $\log M(e_i) = \overline{d_0(e_i,L)} \cdot \log L/e_i$ is effectively a scale average the $\chi^2$ fit results in an {\em average of averages}. For arbitrary dimension distributions this difference in effective definition could be quite significant.

\begin{figure}[ht]
\centering
\includegraphics[width=4in]{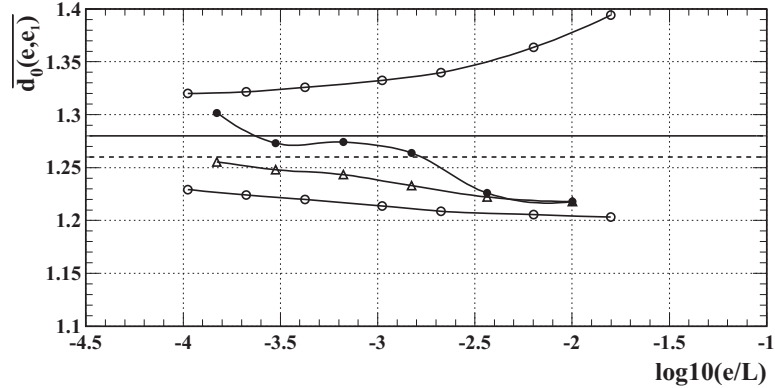}
\caption{Results for limit-based dimensions. Top and bottom open circles are from Eq.~(\ref{single}) for $L'=0.8\, L, 0.4 \,L$ based on corrected box counts from \cite{grasspla1}. Solid dots are from Eq.~(\ref{null}) and open triangles are a running average from Eq.~(\ref{double}), again using the same box counts. The solid horizontal line at 1.28 represents the extrapolation from \cite{grasspla1}, the dashed line at 1.26 represents a box-counting result and Lyapunov upper limit from \cite{russell}}
\label{dimsurv}
\end{figure}

Local averages over six scale intervals were determined from corrected box counts $M(E,e)$ at seven scale points with $\log e  \in [-3.4,-1.6]$ \cite{grasspla1}
\bea \label{null}
D_i &\equiv& \frac{\log M(E,e_{i+1}) - \log M(E,e_i)}{\log L/e_{i+1} - \log L/e_i} = \overline{d_0(E,e_{i+1},e_{i})}  ~~ i \in [1,6].
\eea
These six values are compared (solid points) to other averaging schemes and measurements in Fig. \ref{dimsurv} and to the full scale-local dimension distributions in Fig. \ref{henplotsa1}, where we compare $D_i$ to $\overline{d_0(E,e_{i+1},e_{i})}$ obtained from a scale-local distribution.  Value $d^{\text{b}} = 1.28 \pm 0.01$ was obtained in \cite{grasspla1} by extrapolating the $D_i$ from Eq.~(\ref{null}) to a `zero-scale' limiting value, driving the estimate somewhat high compared to a uniformly-weighted average over the entire $[-4,-1]$ scale interval. Convergence is apparently expected with decreasing scale to a constant value $D \equiv d^{\text{r}}_0$, confusing $\lim_{e \rightarrow 0} \overline{d_q(e,L)}$ with $\lim_{e \rightarrow 0} {d_q(e,\delta e)}$. 

%
%

In Fig.~\ref{dimsurv} and elsewhere the attractor boundary size assumed for the plot axis definition is $L = 2 \cdot 1.8$. Scale values for the solid points ($D_i$ from \cite{grasspla1}) were rescaled accordingly for this figure. Using the $M(E,e_i)$ from \cite{grasspla1} we also plot a variant of Eq.~(\ref{classdim})
\bea \label{single}
D'_i &\equiv& \frac{\log M(E,e_i)}{\log L/e_i - \log L/L'} \approx \overline{d_0(E,e_{i},{L'})} ~~~~~ i \in [1,7],
\eea
where $L'$ is $0.8 \, L$ for the upper open circles and $0.4 \, L$ for the lower open circles. These curves suggest the bias possible from different choices of boundary scale in this definition and the slow convergence in the mean noted in \cite{russell}. With the same $M(E,e_i)$ we also plot (open triangles) the running average.
\bea \label{double}
D''_i &\equiv& \frac{\log M(E,e_i) - \log M(E,e_1)}{\log L/e_i - \log L/e_1} = \overline{d_0(E,e_{i},e_{1})}  ~~~~~ i \in [2,7].
\eea
The solid horizontal line at 1.28 represents the extrapolation from \cite{grasspla1}, the dashed line at 1.26 represents the box-counting result and Lyapunov upper limit from \cite{russell}.

\section{Scale-Local H\'{e}non Dimensions}

We contend that {\em scale-local} R\'{e}nyi dimensions defined over an arbitrary bounded scale interval provide the most fundamental and accessible representation of the dimensionality and correlation structure of an arbitrary distribution, that limit-based dimensions, as asymptotic limits (by extrapolation) of averages, are secondary. For a finite sample with bin contents $n_i(E,e,N)$
\begin{eqnarray} \label{rendim}
d_q(E,e,N) &=& \partial S_q(E,e,N) / \partial \log 1/e \\ \nonumber
&\simeq& {1 \over q-1} {\partial \log \left\langle \sum_{i=1}^{M(E,e,N)} p_i^q(E,e,N) \right \rangle_\phi \over \partial \log e} ,
\end{eqnarray}
where $p_i(E,e,N) \equiv n_i(E,e,N) / N$, and the brackets indicate bin dithering to further reduce bias, especially important near the boundary scale \cite{scanew}. While $M(E,e)$ (and therefore $d_0(E,e)$) can in principle be obtained by correcting void-bin bias as noted above, we do not make a correction in this analysis, relying instead on large sample number $N$ to minimize bias within a specified scale interval.

\subsection{Technical details}

Attractor point-set data are compactly summarized by applying an "atomic" or  microscale binning at some small scale well below the lower limit of the intended working scale interval.  The microbin contents are stored as a zero-suppressed list of bin occupancies.  To calculate the relevant topological measures the microbinning is itself binned at increasing scale (with scale points defined by integer multiples of the microbin scale).  This approach maintains analysis generality while minimizing computation time.

The number of scale points available for analysis (scale resolution) is limited by the microbin scale.  This scale can always be chosen small enough to allow for any requisite scale resolution, at the expense of greater computation time.  In this analysis the microbin scale is chosen to be more than a decade below the lower bound of the working scale interval. This choice insures that integer granularity does not limit the desired scale resolution. A very high resolution study over 10 decades of scale is reported in \cite{sprott}.

Entropy values are calculated for approximately 20 scale points in each decade of scale, implying a scale resolution $\delta \log e = 0.05$, and dimension values are calculated between pairs of neighboring entropy points as ratios of finite differences. Details of scale-local entropy calculation, including bin dithering to minimize boundary-scale bias, are given in \cite{scanew}.

\subsection{General scale-local distribution features}

\begin{figure}[ht]
\centering
\includegraphics[width=4in]{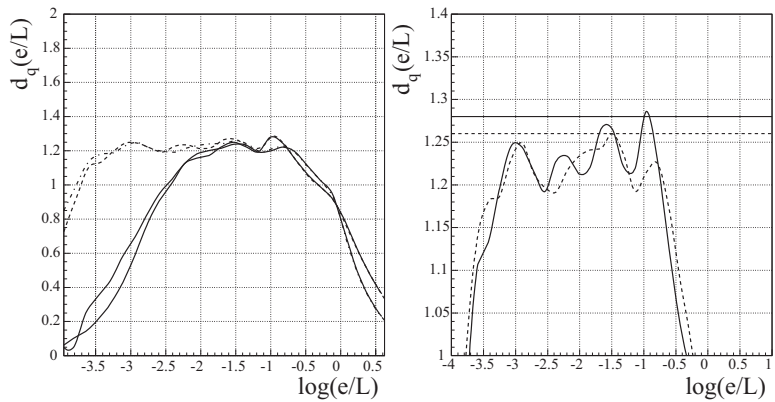}
\caption{Left panel -- dimension distributions for $q = 0,1$ and two sample numbers: N = 10k (solid) and 300k (dashed). Right panel -- N = 300k distributions from left panel for $q = 0$ (solid curve) and $q = 1$ (dotted curve). Expanded vertical scale allows comparison with limit-based values (horizontal lines). \label{HenonDim}}
\end{figure}

In Fig.~\ref{HenonDim} we present scale-local dimensions $d_q(E,e,N)$ for $q = 0$ and $ 1$ (solid and dashed respectively in right panel) for an analysis with scale resolution $\delta \log e \approx 0.15$. Distributions in the left panel rise from zero above the boundary scale (the whole attractor appears as a single point) to values near 1.2 at scales well below the boundary. Distributions return to zero at small scale, reflecting resolution of individual sample points. The right panel with expanded vertical scale indicates that typical variation of either distribution is larger in amplitude than the difference between their means. Horizontal lines at 1.28 (solid) and 1.26 (dashed) reflect estimates of limit-based  dimensions $d_0^{\text{r}}, ~ d_1^{\text{r}}$ respectively in \cite{grassprl,grasspla1}

%
%

\subsection{Scale-local dimension details}

\begin{figure}[ht]
\centering
\includegraphics[width=4in]{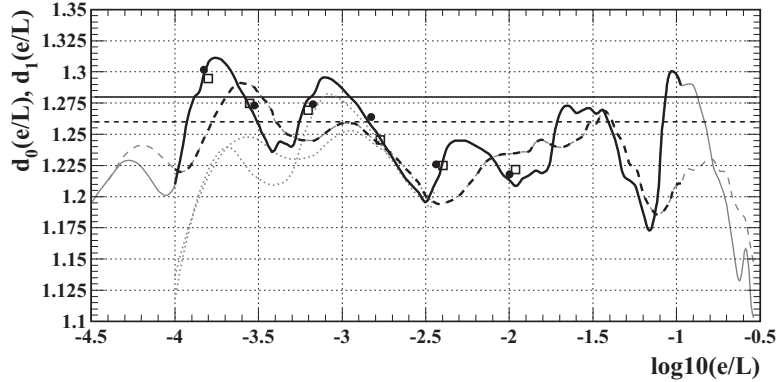}
\caption{Scale-local dimensions for $q=0$ (solid) and $q=1$ (dashed) over four scale decades for 10M map iterations. Dotted curves are for 1M map iterations showing void-bin bias. Solid dots are local scale averages $D_i$ from \cite{grasspla1}. Open boxes are comparable averages of scale-local dimensions showing good agreement. Horizontal lines correspond to limit-based estimates of $d_0^{\text{r}} = 1.28$ \cite{grasspla1} (solid) and of $d^{\text{r}}_1$ and $d^{\text{l}}$ = 1.26 \cite{russell} (dashed).}
\label{henplotsa1}
\end{figure}

Fig.~\ref{henplotsa1} shows higher-resolution distributions of scale-local dimensions $d_0(e)$ (solid) and $d_1(e)$ (dashed) for 10M mapping interactions on scale interval $\log e/L \in [-4,-1]$ also used to form running-average distributions in Fig.~\ref{henplotsa2}. Within this interval the distributions closely approximate those of the parent. Extensions beyond this optimum interval (gray curves) are subject to significant bias (boundary and void-bin). Dotted curves at smaller scale (for 1M iterations) illustrate void-bin bias. Also plotted are the local-average $D_i$ values from \cite{grasspla1} (solid dots) shown in Fig.~\ref{dimsurv}. Averages $\overline{d_0(e_{i+1},e_i)}$ of the scale-local distribution over nearly matching scale intervals (open boxes) are in good agreement.
%
%

The distributions are highly structured (structure in Fig.~\ref{henplotsa1} limited by scale resolution $\delta \log e \simeq 0.05$) and very reproducible (typical variation with different mapping seeds is within two line widths). Scale-local dimension is clearly not point-wise monotonically ordered with $q$ (discussed further in Sec.~\ref{mono}). There is no trend within this scale interval for decreasing structure at smaller scale. It is suggested in another high-resolution study over 11 decades \cite{sprott} that the dimension distribution may itself be self-similar on scale {\em resolution} and does not obviously converge to a limiting value at smaller scale.

\subsection{Running scale averages}

%
\begin{figure}[ht]
\centering
\includegraphics[width=4in]{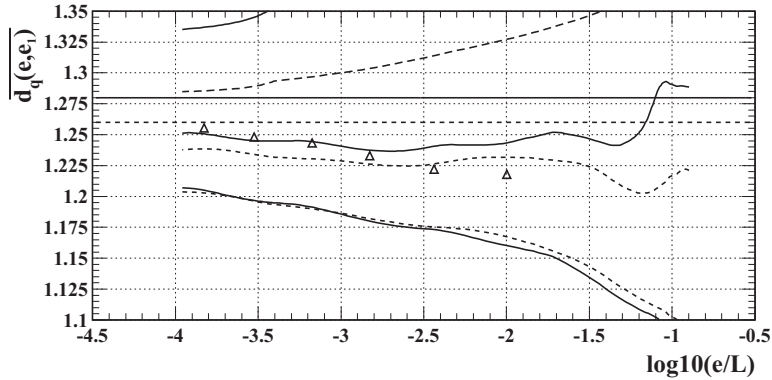}
\caption{Running scale averages $\overline{d_q(e,e_1)}$ for $q = 0,\, 1$ for distributions in Fig.~\ref{henplotsa1} with $\log e_1/L = -0.9$. Open triangles show data from \cite{grasspla1} also as a running average according to Eq.~(\ref{double}). Lower curves show boundary bias due to starting point $\log e_1/L = -0.5$. Uppermost curves show results from modified limit-based definition Eq.~(\ref{single}) and $\log L/L' = 0.1$. Open triangles show data from \cite{grasspla1} as a running average according to Eq.~(\ref{double}).}
\label{henplotsa2}
\end{figure}
%

In Fig.~\ref{henplotsa2} are shown running averages of two scale-local distributions of the form $\overline{d_q(e,e_1)}$ Eqs.~(\ref{dlim},\ref{double}) based on $d_q(e)$ distributions in Fig.~\ref{henplotsa1}. Middle curves (solid: $q=0$, dashed: $q=1$) correspond to $\log e_1/L \approx -0.9$ chosen to minimize bias near the boundary scale. These running averages continued over a semi-infinite scale interval would terminate in limit-based dimension values. The low-scale limiting values in this plot furnish the best {\em estimates} of limit-based dimensions given finite computing resources. Lower curves correspond to $\log e_1/L \approx -0.5$ and are severely biased. Horizontal lines correspond to limit-based estimates of $d_0^{\text{r}} = 1.28$ \cite{grasspla1} (solid) and of $d^{\text{r}}_1$ and $d^{\text{l}}$ = 1.26 \cite{russell} (dashed). Open triangles are data from \cite{grasspla1} treated as a running average according to Eq.~(\ref{double}). Uppermost two curves illustrate Eq.~(\ref{single}) for $\log L/L' \sim 0.1$ showing a different aspect of boundary bias for this dimension definition. Similar results obtain for Eq.~(\ref{avdim}) (there is a small vertical shift due to the different scale weighting).

\subsection{Scale-local dimensions for higher $q$}

%
\begin{figure}[ht]
\centering
\includegraphics[width=4in]{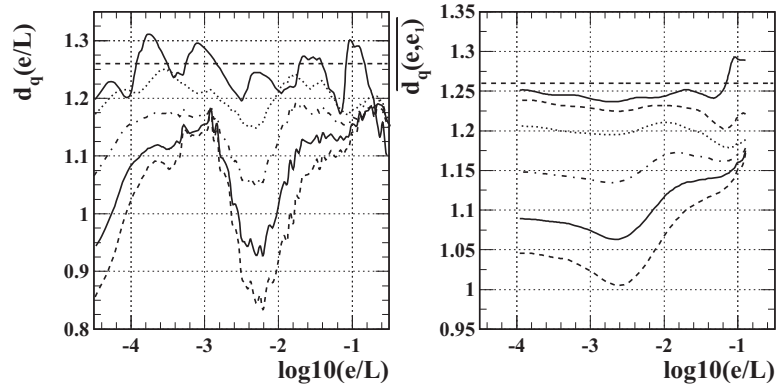}
\caption{Left panel -- scale-local ${d_q(e/L)}$ for $q \in [0,5]$ ($q$ increasing from top (solid) to bottom (dashed) curve). Right panel -- running scale averages $\overline{d_q(e,e_1)}$ of the distributions in the left panel, showing monotonic $q$-ordering for large averaging intervals.}
\label{qplotgen}
\end{figure}
%

The left panel in Fig.~\ref{qplotgen} shows scale-local ${d_q(e)}$ for $q \in [0,5]$ ($q$ increasing from top to bottom curve). We observe large excursions in dimension values (up to 30\% over the scale interval). These results compare well with $D^{eff}(e_i) \equiv \overline{d_q(e_i/2,2e_i)}$ in Fig. 2 of \cite{grasspla3}. Increasing $q$ acts as a contrast adjustment, enhancing sensitivity to large density variations on the attractor. There is a suggestion of periodicity on scale for larger $q$ values. The right panel shows running averages $\overline{d_q(e,L/9)}$, also for $q \in [0,5]$. For averages over a sufficiently large scale interval the running averages are consistent with monotonicity on $q$, as discussed in Sec.~\ref{mono}. Values at the small-scale endpoints are numerical estimators for limit-based dimensions.

\subsection{Monotonicity with $q$} \label{mono}

We observe in Figs.~\ref{HenonDim},~\ref{henplotsa1} that scale-local $d_0$ and $d_1$ are not monotonic with respect to $q$, $q$-ordering alternates with varying scale. Monotonic decrease of R\'{e}nyi {\em entropies} with $q$ has been established for certain conditions. Following \cite{grasspla2}
\bea \label{qderiv}
\partial S_q(e) / \partial q = - \frac{1}{(1-q)^2} \sum_i z_{qi} \log ( z_{qi} / p_i ) ,
\eea
with $z_{qi} \equiv p_i^q / \sum_k p_k^q$.  Since the sum on the right (Kullback information) is non-negative definite the $q$-derivative is never positive. Limit-based dimensions $d_q^{\text{r}} = \lim_{e \rightarrow 0} S_q(e) / \log (L/e)$ then inherit this monotonicity. However, a similar $q$-ordering of scale-local dimensions is not generally valid. The $q$-derivatives for scale-local dimension are
\bea \label{dimq}
\partial d_q(e) / \partial q = \frac{1}{(1-q)^2} \, \frac{\partial}{\partial \log e} \left\{ \sum_i z_{qi} \log ( z_{qi} / p_i ) \right\}.
\eea

There is no {\em a priori} restriction on the sign of the scale derivative on the right (a form of {\em dimension transport} -- see Sec.~\ref{topinv} and \cite{scanew,SCA}). Dimension transport, the scale derivative of scale-local information, can have positive or negative values. However, dimension transport averaged over a scale interval descending from the boundary scale {\em must} be negative-definite, since the integral over that scale interval is the information: $I_q(e,L) = - \int_{\log e}^{\log L} \Delta d_q(e') \, d \log e'$ \cite{scanew}. If the `information' in the bracket of Eq.~(\ref{dimq}) is positive definite (a possible condition on transformations), scale-averaged dimension over the corresponding scale interval is monotone-decreasing with $q$: $\partial \overline{d_q(e,L)} / \partial q \leq 0$. The running averages in the right panel of Fig.~\ref{qplotgen} are indeed monotonic with $q$ over most of the observed scale interval, whereas the scale-local distributions (left panel) are significantly non-monotonic.

\section{Limit-based Dimensions as Topological Invariants} \label{topinv}

The invariance of dimension under a transformation depends on the amount of information generated by the transformation. We show that invariance of limit-based dimensions under transformation holds if the amount of information generated by the transformation is bounded over a semi-infinite scale interval. This condition can be compared to the definition of homeomorphism as a `continuous' or `smooth' bijective map.

We have established that limit-based dimensions are averages of scale-local dimensions over semi-infinite scale intervals. Transformations alter scale-local dimension distributions in the form of {\em dimension transport}, the scale derivative of information \cite{scanew,SCA}.  Some transformations produce transport over a finite scale interval, others  over semi-infinite scale intervals. An example of the latter is the H\'{e}non map itself, which maps a 2D space to a $\sim 1.25$D attractor. Such maps, and recursive transformations like Peano and baker foldings, have no lower-scale boundary: limit-based dimensions are not protected as invariants. 

The change in limit-based R\'{e}nyi dimension with dimension transport resulting from a transformation can be written
\bea \label{diminv}
\Delta d_q^{\text{r}} &=& \lim_{e \rightarrow 0} \left\{ \int_{\log e}^{\log L} \Delta d_q(e') d \log e' / \int_{\log e}^{\log L}  d \log e' \right\} \\ \nonumber
&=&\lim_{e \rightarrow 0}  \overline{\Delta d_q(e,L)},
\eea
where {\em e.g.} dimension transport for a 2D embedding space is a signed number bounded by $\Delta d_q(e) \in [-2,2]$. If a transformation generates dimension transport over a {\em finite} scale interval its integral in Eq.~(\ref{diminv}), which is information $ I_q(e,L)$, should be zero in the asymptotic limit, and limit-based dimensions remain invariant. 

The scale dependence of dimension transport reflects the scale structure of transformations and thus provides a basis for classifying them. An analogy can be drawn between the logarithmic system of entropy, information, dimension and dimension transport which describes correlation over arbitrary scale intervals and the linear system of correlation integral, cumulant, autocorrelation and autocorrelation difference suitable for correlation structure restricted to small scale intervals. Difference quantities which compare two distributions (possibly related by a transformation) can be defined in the two systems by
\bea
\Delta C_q(e) &=& \int^e_0 \Delta A_q(e') \, d e' \\ \nonumber
\Delta \log C_q(e) &=& (1-q) \int_{\log e}^{\log L} \Delta d_q(e') \, d \log e',
\eea
where $\Delta A_q(e)$ is an autocorrelation difference or {\em net} autocorrelation, $\Delta d_q(e) = - \partial I_q(e) / \partial \log e$ is dimension transport, $\Delta C_q(e)$ can be related to cumulants and $\Delta \log C_q(e)$ to information. Autocorrelation and scale-local dimension play analogous roles as local or differential correlation measures on scale, in linear and logarithmic systems respectively. Entropies and correlation integrals are the corresponding integral quantities. 

Changes in correlation result in {\em transport} of autocorrelation or dimension on scale as conserved quantities. Choice of linear or logarithmic correlation measures depends on the scale structure of distributions. The linear system (with a long history of development) is better suited for linear periodic systems; the logarithmic system for complex, nonlinear aperiodic systems.

\section{Conclusions}

Topology seeks static properties of transformation-invariant sets amenable to rigorous treatment. Science seeks to describe the evolution of transient complex systems viewed imperfectly. Science thus requires a broader descriptive system, a generalized, self-consistent treatment containing topological invariance and limit-based measures as a special case. Studies of nonlinear dynamics have provided a fruitful interface between mathematical idealization and real-world arbitrariness. After a period of rapid development one could pursue a path toward greater rigor which emphasizes a subset of interesting dynamics and methods and favors limit-based measures. Alternatively, one could strive to encompass arbitrarily scale-dependent systems in a generalized scale-local theory of dimension, with limit-based dimension as a special case.

In this paper we compare scale-local and limit-based dimensions through analysis of a well-known distribution -- a strange attractor of the H\'enon map. We find that limit-based dimension is an asymptotic scale average of scale-local dimension. The former is not $\lim_{e \rightarrow 0} d_q(\log e)$, a limiting scale-local value at zero scale, but rather $\lim_{e \rightarrow 0} \overline{d_q(\log e,\log L)}$, an average over a semi-infinite scale interval. The general quantity is $\overline{d_q(\log e_1,\log e_2)}$, an average over arbitrary scale interval with limiting cases $\lim_{e \rightarrow 0} \overline{d_q(\log e_,\log L)}$ (limit-based dimension) and $\overline{d_q(\log e - 0.5 \, \delta \log e,\log e + 0.5 \, \delta \log e)}$ (scale-local dimension at scale $\log e$ and resolution $\delta \log e$). Running average $\overline{d_q(\log e,\log L)}$ descending from a boundary scale {\em estimates} limit-based dimension in experimental or computational contexts.

Scale-local R\'enyi dimensions of the H\'{e}non map attractor analyzed here are {\em highly scale dependent}, with variations on scale up to 30\% of mean values, no indication of convergence to a fixed limit and no consistent monotonicity on $q$ index. Running averages based on scale-local distributions do exhibit monotonicity on $q$ index for significant averaging intervals. Running scale averages provide the best means for estimating limit-based dimension values from numerical or experimental data, but as statistical estimators these averages are subject to several significant sources of bias, as demonstrated in this and other analyses. 

We re-express dimension invariance in terms of scale averages, information and dimension transport. Limit-based dimension defined on a semi-infinite scale interval is invariant under transformations corresponding to {\em finite information}, or equivalently dimension transport restricted to a finite scale interval (recursive transformations for example do not satisfy this condition). By the same argument, dimension {\em estimators} derived from finite scale intervals are not generally invariant under transformations, in fact may be used to study the scale-dependent structure of transformations. Dimension transport provides an alternative classification basis.


\end{document}